\documentclass[twocolumn,aps,prl,showpacs]{revtex4}

\usepackage{graphicx}% Include figure files
\usepackage{subeqnarray,amsmath}% Number equations with letter
\usepackage{dcolumn}% Align table columns on decimal point
\usepackage{bm}% bold math

\begin{document}

\preprint{APS/123-QED}

\title{Gating of a molecular transistor: Electrostatic and Conformational}

\author{Avik W. Ghosh, Titash Rakshit and Supriyo Datta}
\affiliation{School of Electrical and Computer Engineering, Purdue
University, W. Lafayette, IN 47907}%

\date{\today}% It is always \today, today,

\begin{abstract}
We derive a general result that can be used to evaluate and compare
the transconductance of different field-effect mechanisms in molecular
transistors, both electrostatic and conformational. The electrostatic
component leads to the well-known thermal limit in the absence of
tunneling.  We show that in a standard three-terminal geometry and in
the absence of strong electron-phonon coupling, the conformational
component can lead to significant advantages only if the molecular
dipole moment $\mu$ is comparable to $et_{\rm{ox}}$, $t_{\rm{ox}}$
being the thickness of the oxide. Surprisingly this conclusion is
independent of the ``softness'' of the conformational modes involved,
or other geometrical factors. Detailed numerical results for specific
examples are presented in support of the analytical results.
\end{abstract}
\bigskip

%31.15.Ar Ab initio calculations
%81.07.Nb Molecular Nanostructures
%81.07.Lk Nanocontacts
%85.65.+h Molecular electronic devices
%72.10.Bg General formulation of transport theory
%72.20.Dp General theory, scattering mechanisms of conductivity
%73.23.-b Electronic transport in mesoscopic systems
%73.40.Sx Metal-semiconductor-metal structures
%73.63.-b Electronic transport in mesoscopic or nanoscale materials and structures
%2col
%end of wide text
% Avik check PACS from PRL.

\maketitle
%2col

\begin{figure} [ht]
\vspace{3.4in}
\hskip -7.3cm{\includegraphics{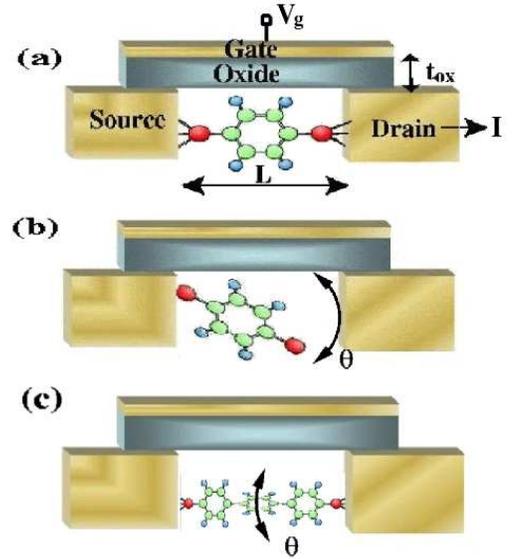}}
\vskip -1.5cm
\caption{In a standard transistor (a) the gate voltage $V_g$ controls
the current $I$ by controlling the induced charge. But in conformational
transistors the gate controls the current through the configurational
parameter $\theta$. This could be either (b) through a variation in
coupling with the contact, or (c) a variation in transmission through
the molecule itself.}
\label{f0}
\end{figure}

Device miniaturization is progressively heading towards solid state
electronic components that are molecular in nature
\cite{rReed,rMolel}.  Molecular transistors are of great current
interest from both basic and applied points of view.  Despite
theoretical proposals \cite{rlangemberly,rWada,rEllenbogen} and
experimental reports \cite{rMcEuen,rJoachim,rKagan} of three-terminal
molecular devices, their general physical principles are not yet well
understood. In a standard silicon MOSFET the gate modulates the
current by controlling the channel charge through its electrostatic
potential. Good transistor action requires that this channel
potential respond much more to the gate than to the drain, implying
an oxide thickness $t_{\rm{ox}}$ that is much smaller than the channel
length $L$ (Fig.~\ref{f0}a). A molecular transistor operating on
electrostatic principles ought to have the same design limitation, so
that even nominal gate control in a 10 \AA~molecule such as phenyl
dithiol (PDT) \cite{rReed} demands an oxide that is prohibitively thin
\cite{rFET}.  It is therefore natural to investigate alternate
principles of molecular transistor action, such as by utilizing
conformational degrees of freedom \cite{rWada}. For instance, a gate
field coupled with a molecular dipole moment $\vec{\mu}$ could cut
off the current by tilting it away from a contact (Fig.~\ref{f0}b),
or by twisting one part relative to the other, breaking its conjugation
\cite{rDinitro,rsamanta} (Fig.~\ref{f0}c).  Such a mechanism could respond
more strongly to the gate than the drain if $\vec{\mu}$ is engineered
to lie along the source-drain field.

In this paper, we derive a {\it{general expression}} for the
transconductance per unit current $g_m/I = (1/I)(\partial I/\partial
V_g)$ ($I$: current, $V_g$: gate voltage) \cite{rfom} for different
transistor mechanisms, both electrostatic and conformational \cite{rset}. For a
large separation of time scales between electronic and vibronic modes,
as is typical in room temperature molecular conductors \cite{rPulay}, the
electrostatic and conformational transconductances are {\it{additive}},
with a well-known electrostatic maximum:
\begin{equation}
\left[ g_m^{\rm{es}}/I\right]_{\rm{max}} = e\beta = (25 mV)^{-1} = 40 V^{-1}
\label{e0a}
\end{equation}
($\beta$: inverse thermal energy), achieved only if $t_{\rm{ox}}
\ll L$ {\it{and if there is negligible
tailing in the density of states}}. In addition, we find a fundamental
conformational maximum, independent of modal stiffness or molecular
geometry:
\begin{equation}
\left[ g_m^{\rm{conf}}/I\right]_{\rm{max}} = e\beta ({{\mu}/{et_{\rm{ox}}}}).
\label{e0b}
\end{equation}
These simple expressions for gate control are
later corroborated with detailed numerical results that self-consistently
combine a semi-empirical description of the device Hamiltonian with a
nonequilibrium Green's function (NEGF) description of transport. Based
on these results, it seems that good molecular transistor action in a 
standard MOS geometry with a modest gate insulator thickness
would require degenerately doped semiconducting contacts, along with the
incorporation of a molecular dipole dipole aligned along the source-drain
direction and large enough to overcome any room-temperature thermal
averaging over conformations, ie, larger than $et_{\rm{ox}}$.

We assume above that conducting and non-conducting molecular states 
lie in the same valley in the conformational potential
landscape. The gate voltage shifts the minimum within this valley
but thermal excitations lead to a statistical average over the entire
valley, making it difficult to change the current any faster than the
fundamental limit implied by Eq.~\ref{e0b}. One could get  around this
limit if the conducting and non-conducting molecular states belong
to distinct metastable valleys \cite{rWada}, such that thermal averaging
takes place only over one valley or the other.  The role of the gate
then is to shift the molecule from one valley to the other through a large
impulsive force. Such a mechanism, however, is very different from the
operating principles of present day MOS devices.

We will now discuss these issues quantitatively by deriving Eq.~\ref{e2}
for $g_m/I$ and evaluating its electrostatic and conformational components
for specific illustrative examples, with detailed numerical calculations
for support.

{\it{Transconductance equation}}.
A source-drain bias $V_d$ splits the contact chemical
potentials $\mu_{1,2}$ by the applied voltage.  The
molecular I-V is obtained from Landauer theory \cite{rDattabook}, which
can be recast in the less familiar form:
\begin{eqnarray} 
I &=& {{2e}\over{h}}\int_{\mu_1}^{\mu_2} dE ~\langle ~\tilde{T}(E)~\rangle;
\nonumber\\ 
\tilde{T}(E) &=& T(E) \otimes F_T(E).
\label{e00}
\end{eqnarray} 
Thermal broadening appears \cite{rBagwell} as a convolution
$\otimes$ of the transmission $T(E)$ at {\it{zero lead temperature}} with
the thermal broadening function $F_T(E)=$ $-\partial f(E)/\partial E$,
$f(E) =$ $1/[\exp(E\beta)+1]$.  $\langle \ldots\rangle$ gives a thermal
average over various molecular configurations $\{x\}$ with probabilities
set by the conformational potential energies $U(\{x\})$:
\begin{equation}
\langle \tilde{T}(E)\rangle = {{\sum_i \tilde{T}(E;x_i)w_i}\over{\sum_i w_i}},~~~ w_i = \exp{\left[-\beta U(x_i)\right]},
\label{e0}
\end{equation}
$x_i$ denoting the coordinate of the ith configuration.  From
Eq.~\ref{e00}, $I = (2e/h)(\mu_1 - \mu_2)\langle \tilde{T}_0\rangle$,
$\tilde{T}_0$ being the {\it{average}} value of $\tilde{T}(E)$ over
the energy range $\mu_1 < E < \mu_2$. 
From Eq.~\ref{e0}, straightforward algebra gives $g_m = g_m^{\rm{es}} +
g_m^{\rm{conf}}$, where
\begin{subeqnarray}
\slabel{e2a}
{{g_m^{es}}\over{I}} &=& {{1}\over{\langle \tilde{T}_0\rangle}}\left\langle {{\partial \tilde{T}_0}\over{\partial V_g}}\right\rangle,\\
\slabel{e2b}
{{g_m^{conf}}\over{I}} &=& {{\beta}\over{\langle \tilde{T}_0\rangle}}\Biggl[\langle \tilde{T}_0\rangle \left\langle {{\partial U}\over{\partial V_g}}\right\rangle - \left\langle \tilde{T}_0{{\partial U}\over{\partial V_g}}\right\rangle \Biggr].
\label{e2}
\end{subeqnarray}
The transconductance has two additive contributions: an
{\it{electrostatic}} term describing how the gate controls the channel
charge, and a {\it{conformational}} term describing how the gate
controls the transmission by deforming the molecule \cite{rPulay}. We
now discuss these terms one by one.

{\it{Electrostatic control}}.  Electrostatically the gate modulates
the self-consistent potential, effectively moving the
molecular energy levels relative to the contacts. A complete treatment
of this effect requires a self-consistent solution of the Schr\"odinger-Poisson
equations.
However, one can get a qualitative description by
assuming that a gate bias rigidly shifts the energy levels and
hence the transmission: $T(E,V_g) \approx T(E-\gamma eV_g)$. The
parameter $\gamma$ tells us the average molecular potential $V_m$
in response to a change in gate bias: $\gamma
\equiv \partial V_m/\partial V_g$. Using Eq.~\ref{e2a}, 
\begin{equation}
{{g_m^{\rm{es}}}\over{I}} = {{e\gamma}\over{\langle\tilde{T}_0\rangle}} \left\langle {{\partial \tilde{T}_0}\over{\partial E}}\right \rangle .
\label{e2.5}
\end{equation}
$\gamma \sim C_G/(C_G + C_S + C_D + C_Q)$, where $C_{G,S,D}$
represent the effective capacitances coupling the molecule to the
gate, source and drain respectively, while $C_Q$ is the quantum 
capacitance proportional to the molecular density of states 
\cite{rQC}.  One could view $C_{G,S,D}$
as a representation of Poisson's equation, and $C_Q$ as a linearized
representation of Schr\"odinger's equation.  Under OFF conditions $C_Q$
is negligible and $\gamma$ can be made close to one in a well-designed
FET by making $t_{\rm{ox}} \ll L$, so that $C_G \gg C_{S,D}$. However,
as we noted earlier, this becomes increasingly difficult as we try
to engineer ultrasmall devices that are only tens of atoms long ($L
\sim 1-5$ nm). The electrostatic parameter $\gamma$ is then severely
reduced, motivating us to look for non-electrostatic mechanisms for
control. Note that the electrostatic restriction on $\gamma$ is 
fundamental, and may not be easy to handle without using large dielectric 
constant insulators with a small effective thickness.
% insulating cages into the molecule itself \cite{rMOSES}.

Eq.~\ref{e2.5} also elucidates the role of
a sharp transition in $\tilde{T}_0$ as a function of $E$ in realizing
good electrostatic control. A metallic conductor with constant $\tilde{T}_0(E)$
cannot be used to build an electrostatic switch. Semiconductors by contrast have
a band-edge where $\tilde{T}_0(E)$ drops to zero over a very
small energy range $\Delta E$, so that from Eq.~\ref{e2.5},
\begin{equation}
{{g_m^{\rm{es}}}/{I}} \approx {{e\gamma}/{\Delta E}}.
\end{equation}
Since $\tilde{T}_0(E)$ represents a convolution
of $T(E)$ and $F_T(E)$, the minimum value of $\Delta E$ equals that
for $F_T(E)$, which is $\sim 1/\beta$. This leads
to the well-known upper limit of $e\beta$ for $g_m^{\rm{es}}/I$, corresponding
to $\gamma = 1$ and $\Delta E = 1/\beta$. An important point to note 
is that increased tunneling in nanodevices can make $T(E)$ 
non-zero below bandgap, with $\Delta E \gg 1/\beta$. This condition
is made worse by the use of metallic (rather than semiconducting) source and
drain regions which is common for molecular devices \cite{rReed,rMolel}. 
In view of recent progress in growing molecules on 
silicon surfaces \cite{rWolkow}, using {\it{doped semiconducting contacts}} seems a realizable 
and highly desirable ideal.

{\it{Conformational control}}.  The conformational mechanism (Eq.~\ref{e2b})
operates by changing the relative energies $U$ for different
molecular configurations $\{x_i\}$. As expected, we get zero
transconductance if  $\tilde{T}_0$
is independent of the configuration $\{x_i\}$, or its variation
is uncorrelated with $\partial U/\partial V_g$.
But if $\tilde{T}_0$ and $\partial U/\partial V_g$ are {\it{negatively
correlated}} such that configurations with larger transmission
$\tilde{T}_0$ have their energies $U$ reduced by the gate voltage
$V_g$ (and hence made more likely) then the current increases,
with a positive $g_m^{\rm{conf}}/I$. 
To estimate $g_m^{\rm{conf}}/I$, we write:
\begin{equation}
U(\theta;V_d;V_g) = U_0(\theta) - (\mu V_g/t_{\rm{ox}})\sin{\theta}-(\mu V_d/L)\cos{\theta},
\label{e8}
\end{equation}
where $\theta$ is the angle between the molecular dipole $\vec{\mu}$ and the 
source-drain field $V_d/L$. Aligning the dipole along the latter
eliminates its torque \cite{rsolomon}, giving the gate field $V_g/t_{\rm{ox}}$
an obvious superiority. Using Eq.~\ref{e8} in Eq.~\ref{e2b},
\begin{equation}
{{g_m^{\rm{conf}}}\over{I}} =  -\beta{{\mu}\over{t_{\rm{ox}}}}
\Biggl[{{ \langle \tilde{T}_0\rangle\langle\sin{\theta}\rangle - \langle \tilde{T}_0\sin{\theta}
\rangle }\over{\langle \tilde{T}_0\rangle}} \Biggr].
\label{e3.5}
\end{equation}
The quantity within parantheses has a maximum value of one,
indicating that the maximum conformational transconductance is given by
$\mu\beta/t_{\rm{ox}}$ as stated earlier (see Eq.~\ref{e0b}). If $\mu$
is comparable to $et_{\rm{ox}}$, then it is indeed possible for this
mechanism to provide respectable levels of control.  For a 10 \AA~oxide,
$\mu/et_{\rm{ox}}\sim$ 0.15 for an aromatic molecule with one redox NO$_2$
sidegroup per benzene ring. Although it is hard to squeeze in any more
dipolar subgroups per ring, one could incorporate large effective dipoles
external to the molecule, using a strong piezoelectric gate-molecular
coupling, for instance \cite{rJoachim}.  In any case, a conformational
transconductance that is, say, a tenth of $e\beta$ may still be useful
in view of the difficulties with the electrostatic mechanism discussed
earlier. 

Significantly, the upper limit on the conformational
transconductance (Eq.~\ref{e0b}) is unaffected by the `softness' of
the modes described by $U_0(\theta)$ in Eq.~\ref{e8}. For the bending
and twisting configurations in Figs.\ref{f0}(b,c), we write:
\begin{eqnarray}
U_0(\theta) &=& U_0^{\rm{bend}}\left(\theta - \theta_0\right)^2/2,\nonumber\\
U_0(\theta) &=& U_0^{\rm{twist}}\left(1 - \cos{2\theta}\right)/2.
\end{eqnarray}
Twisting a molecular bond is much easier than bending it
($U_0^{\rm{bend}}\sim 2$ eV \cite{rlangm}, $U_0^{\rm{twist}}\sim 0.03$
eV \cite{rtolane});
however, twisting also makes thermal averaging more significant, so that no
advantage is gained as far as $g_m$ is concerned.

The quantity inside parentheses in Eq.~\ref{e3.5}
can ideally have a maximum value of one if $\tilde{T}_0$ is very sharply varying.
In practice it is smaller,
depending on the variation of $\tilde{T}_0$ with $\theta$. As an
example, we model a self-assembled PDT monolayer (Fig.~\ref{f0}b) 
chemisorbed on a Au(111) substrate,
varying the tilt angle from default ($\theta_0 \sim$ 20$^0$) to upright 
position and increasing the sulphur-gold coupling
exponentially. Although accidental symmetries at specific angles of
the highly directional {\it{orbital}} wavefunctions quench the overlap
\cite{rPavel,rsamanta}, such orientational effects wash
out on averaging over various possible positions of the sulphur
and gold surface atoms in the upright configuration. The dominant
angular dependence in $T_0$ then arises from
the overlap between sulphur and gold {\it{radial}} wavefunctions:
$T_0(\theta) \propto \exp{\left[-2ZL(1-\cos{\theta})/na_0\right]}$, 
($Z = 3$: screened nuclear charge, $n$: principal quantum number of sulphur,
$a_0=0.529$\AA).
This gives a correlation in
Eq.~\ref{e3.5} of $\sim$ 0.1 at room temperature ($\mu \approx 8$
Debye,$V_g/t_{\rm{ox}} = 1V/nm$). 
This variation can be made to approach a delta-function by
increasing $L$, giving a stronger correlation. 
Similarly for the rotational configuration in 
Fig.~\ref{f0}c we find $T_0(\theta) \propto \cos^4\theta$
\cite{rMalus}, leading to a configurational average $\sim$ 0.3. The correlation
depends on the variation $T_0(\theta)$, but also on the weighting factors
which in turn depend on the bond stiffness, temperature, gate field and dipole
moment.

\begin{figure}[ht]
\vspace{2.6in}
{\includegraphics{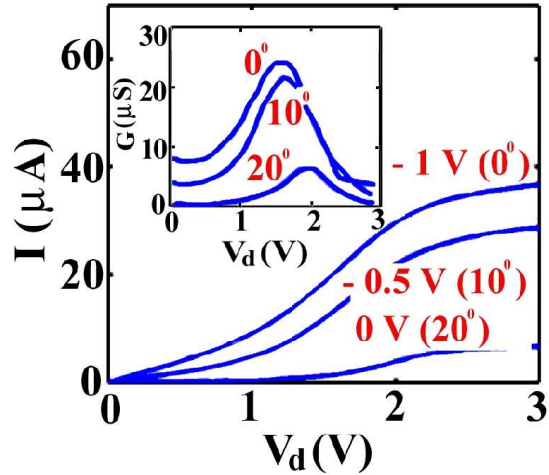}}
\vskip -0.1in
\caption{ The I-V for PDT at a given
angle saturates on crossing the highest occupied molecular
orbital (HOMO) level, increasing with decreasing tilt angle.
The electrostatic gate influence (inset) causes a
shift in the conductance peak towards lower voltages for p-type
(HOMO) conduction.}
\label{f1}
\end{figure}

{\it{Numerical Results.}} We now present detailed calculations of the 
I-V for the structures in Figs.~\ref{f0}b,\ref{f0}c.
The transmission at a given angle is calculated using the nonequilibrium
Green's function (NEGF) formalism
\cite{rabinitio} with an extended H\"uckel Hamiltonian $H$. We 
solve
\begin{eqnarray} G(E) &=& \left(
ES - H - U(\rho) - \Sigma_1 - \Sigma_2\right)^{-1}, \nonumber\\
\rho &=& \int_{-\infty}^\infty \left(f_1G\Gamma_1 G^\dagger +
f_2G\Gamma_2G^\dagger\right) dE/2\pi, \nonumber\\
T &=& {\rm{trace}}\left(\Gamma_1G\Gamma_2G^\dagger\right),
\label{eT}
\end{eqnarray} 
self-consistently ($S$: overlap matrix, $f_{1,2}= f(E-\mu_{1,2})$).
An ideal Au(111) surface geometry \cite{rabinitio} is used to 
calculate the contact self-energies $\Sigma_{1,2}$ and level broadenings
$\Gamma_{1,2}$.  The self-consistent potential $U(\rho)$ is obtained by
solving Laplace's equation for the device geometry with a 10 \AA~
oxide and a Hubbard-type electron-electron term.

Fig.~\ref{f1} shows the zero-temperature I-V for a PDT relay for various
tilt angles. The current tends to saturate on crossing the
HOMO level, and increases with decreasing tilt (increasing $V_g$).
In addition, there is a conventional electrostatic gate influence
\cite{rFET}.  For a p-type (HOMO) conduction, a negative gate bias
raises the molecular levels relative to the contacts, producing a
{\it{lateral shift}} in the conductance ($G=\partial I/\partial V_d$)
towards lower $|V_d|$ (inset).

\begin{figure}[ht]
\vspace{1.6in}
{\includegraphics{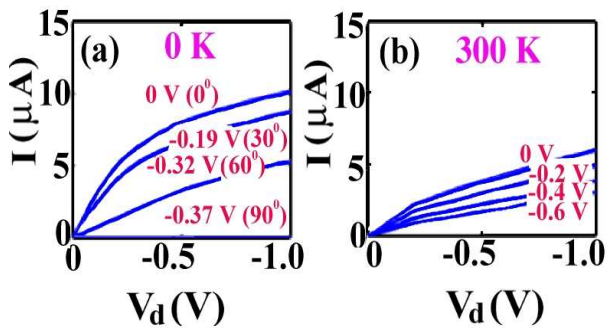}}
\vskip -0.3cm
\caption{I-V of the Tour-Reed molecule \protect\cite{rTourReed}
for various gate voltages at (a) 0 K and (b) 300 K,
$E_F$ lying in the tail of the HOMO level.
At 300 K there is some gate modulation, although it is hard to
completely switch the current off.}
\label{f5}
\end{figure}

Fig.~\ref{f5} shows the three-terminal I-V of the Tour-Reed molecule
\cite{rTourReed} containing a nitroamine redox group at (a) 0 K and
(b) 300 K \cite{rndr}.  $E_F$ is assumed to lie in the HOMO tail,
giving an ohmic I-V at low $V_d$. The impedance is compromised by
the poor electrostatics due to the oxide thickness and MIGS, as
discussed earlier. The low-temperature $g_m/I$ is impressive due to
the low torsion constant and the large dipole, a mere 400 mV gate bias
reducing the current two hundred times. At room temperature, however, the
molecule samples a wide range of angles, reducing $g_m/I$
substantially. The gate-modulation continues to be observable, but the
current is a lot harder to switch off. A lower torsion constant gives
better gate control, but increases thermal effects as well.

The performance of a transistor also depends on its operating speed.
While complicated isomeric cis-trans rotations tend to be slow,
bond rotations are much faster ($\sim$ 10-100 GHz), and need to be
damped out without consuming too much power. Image forces and Van der
Waals interactions with electrodes \cite{rAlluru} as well as steric and
hydrogen bonding interactions between molecules \cite{rEllenbogen} in a
mixed monolayer could damp out such oscillations. 

We have shown that conformational transitions can aid electrostatic gate
control significantly if we could engineer a large molecular dipole
along a suitable direction. Ways around such design restrictions
require going beyond Eq.~\ref{e8}, employing other kinds of
electromechanical gate-molecular coupling (piezoelectric for example),
or using non-traditional, bistable potentials with large impulsive
gate voltages \cite{rWada}.  Eq.~\ref{e2} gives us a general way to 
quantitively compare the transconductance of these various field-effect 
mechanisms in molecular transistors.

We thank P. Damle, D. Monroe, P. Solomon and M. Lundstrom for useful
discussions, and T. Raza for the image in the table of
contents. This work is supported by the US ARO and the Semiconductor
Technology Focus Center on Materials, Structures and Devices.


\begin{thebibliography}{100}
\bibitem{rfom} We use $g_m/I$ rather than $g_m$ to ensure that this `figure of
merit' is a measure of the effectiveness of the mechanism rather than of the number
of molecules involved.
\bibitem{rset} We are not considering resonant or single-electron tunneling
devices. The former can be analyzed using Eqs.~\ref{e00},\ref{e0},\ref{e2}, 
but the latter needs a many-particle description.
\bibitem{rsolomon} Eliminating source-drain torque may require a double-gated
device (P. Solomon, private communications).
\bibitem{rndr} Experimentally these molecules show negative differential 
resistance \cite{rTourReed}, but the mechanism is unclear and does not follow
from the formulation used here.
\end{thebibliography}
\end{document}